\documentclass{ws-procs10x7}

\begin{document}

\title{Precision measurements from the NOMAD experiment}

\author{R. Petti}

\address{CERN, CH-1211 Gen\'eve 23,
Switzerland\\E-mail: Roberto.Petti@cern.ch \\
(for the NOMAD collaboration)}

\twocolumn[\maketitle\abstract{
The NOMAD experiment collected unprecedent neutrino data samples, 
matching both the large statistics of massive calorimeters and 
the reconstruction quality of bubble chambers. This paper describes 
the determination of the weak mixing angle which is ongoing in NOMAD, 
with a target precision of $\sim 1\%$. In addition, measurements 
of the $\nu_{\mu}$ quasi-elastic cross-section and of neutrino 
Charged Current differential cross-section on carbon are presented.
}]

\section{Introduction}
\label{sec:intro} 

The NOMAD experiment was designed to search for $\nu_{\tau}$
appearance from neutrino oscil\-la\-tions in the CERN wide-band
neutrino beam produced by the 450 GeV proton synchrotron.
%The detection of an oscillation signal in NOMAD relied on the 
%identification of the $\tau$ lepton from $\nu_{\tau}$ charged-current 
%(CC) interactions using kinematic criteria. 
The single-particle reconstruction and lepton identification capability 
of the NOMAD detector allowed the search for $\nu_{\tau}$ appearance in most 
of the leptonic and hadronic $\tau$ decay channels~\cite{nmnt} and 
also to look for $\nu_{\mu} \rightarrow \nu_{e}$ oscillations~\cite{nmne}. 
No evidence for oscillations was found.  

A second phase of the NOMAD analysis started after the completion 
of the oscillation searches, with the aim of exploiting the high 
quality of the available neutrino data samples for precise 
measurements of cross-sections and particle production. 
This activity could benefit from the beam and detector 
studies performed for the oscillation searches.

\section{Detector and data samples} 
\label{sec:detdata} 

The NOMAD detector is described
in detail in Ref.~\cite{NOMADNIM}. Inside
a 0.4 T magnetic field there is an active target consisting of
drift chambers (DC)~\cite{DC} with a fiducial mass of about 2.7 tons
and a low average density (0.1 g/cm$^3$). The main target, 405 cm
long and corresponding to about one radiation length,
is followed by a transition radiation detector (TRD)~\cite{TRD}
for electron identification, a preshower detector (PRS), and
a high resolution lead-glass electromagnetic
calorimeter (ECAL)~\cite{ECAL}. A hadron calorimeter (HCAL) and
two stations of drift chambers for muon detection are located
just after the downstream part of the magnet coil. An iron-scintillator 
sampling calorimeter with a fiducial mass of about 17$t$ (FCAL) is 
located upstream of the central part of the NOMAD target. 
The detector is designed to identify leptons and to measure
muons, pions, electrons and photons with comparable resolutions.
Momenta are measured in the DC with a resolution:
$$
\frac{\sigma_p}{p}\simeq \frac{0.05}{\sqrt{L[m]}}\oplus
\frac{0.008 \times p[GeV/c]}{\sqrt{L[m]^5}}
$$
where L is the track length and $p$ is the momentum.
The energy of electromagnetic showers, $E$, is measured
in the ECAL with a resolution:
$$
\frac{\sigma_E}{E}=0.01 \oplus \frac{0.032}{\sqrt{E[GeV]}}.
$$

The relative composition 
of CC events in NOMAD is estimated~\cite{beam} to be 
$\nu_{\mu}$,:\,$\bar{\nu}_{\mu}$,:\,$\nu_e$,:\,$\bar{\nu}_e$ =
1.00\,: \,0.0227\,: \,0.0154: \,0.0016, with average
neutrino energies of 45.4, 40.8, 57.5, and 51.5 GeV, respectively.
Neutrinos are produced at an average distance of 625 m from the detector.

The NOMAD experiment collected data from 1995 to 1998. Most of the running, 
for a total exposure of $5.1\times 10^{19}$ protons on target (pot), was 
in neutrino mode. This resulted in three distinct data samples, according 
to the different targets: $1.3\times 10^{6} \nu_{\mu}$ CC interactions 
from the drift chambers (mainly carbon), $1.5\times 10^{6} \nu_{\mu}$ CC interactions 
from the region of the magnet coil (mainly aluminium) located in front of the DC 
and $1.2\times 10^{7} \nu_{\mu}$ CC interactions from FCAL (iron).

\section{Precision measurements} 
\label{sec:prec} 

\begin{figure}%1
\epsfxsize200pt
\figurebox{120pt}{160pt}{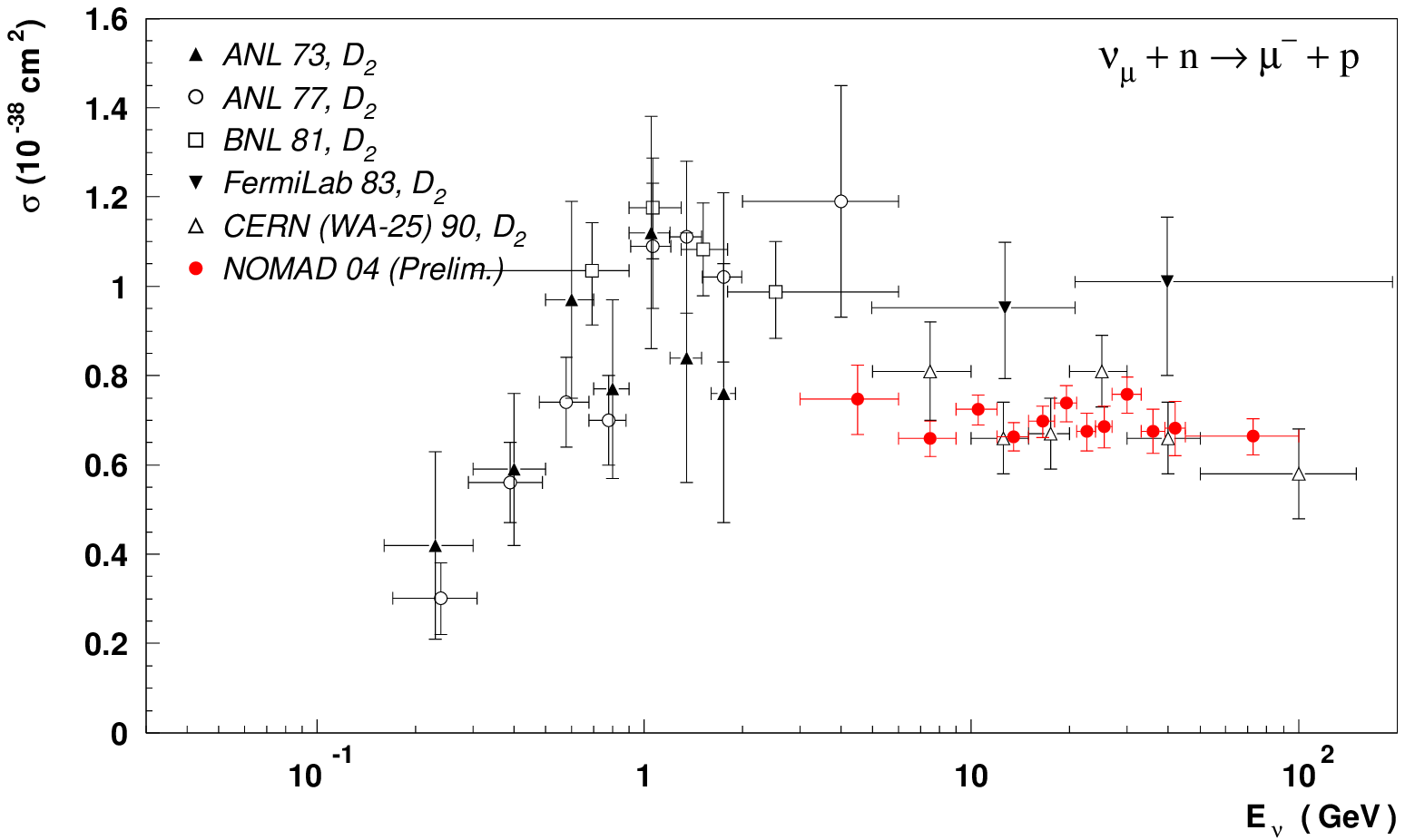}
\epsfxsize200pt
\figurebox{120pt}{160pt}{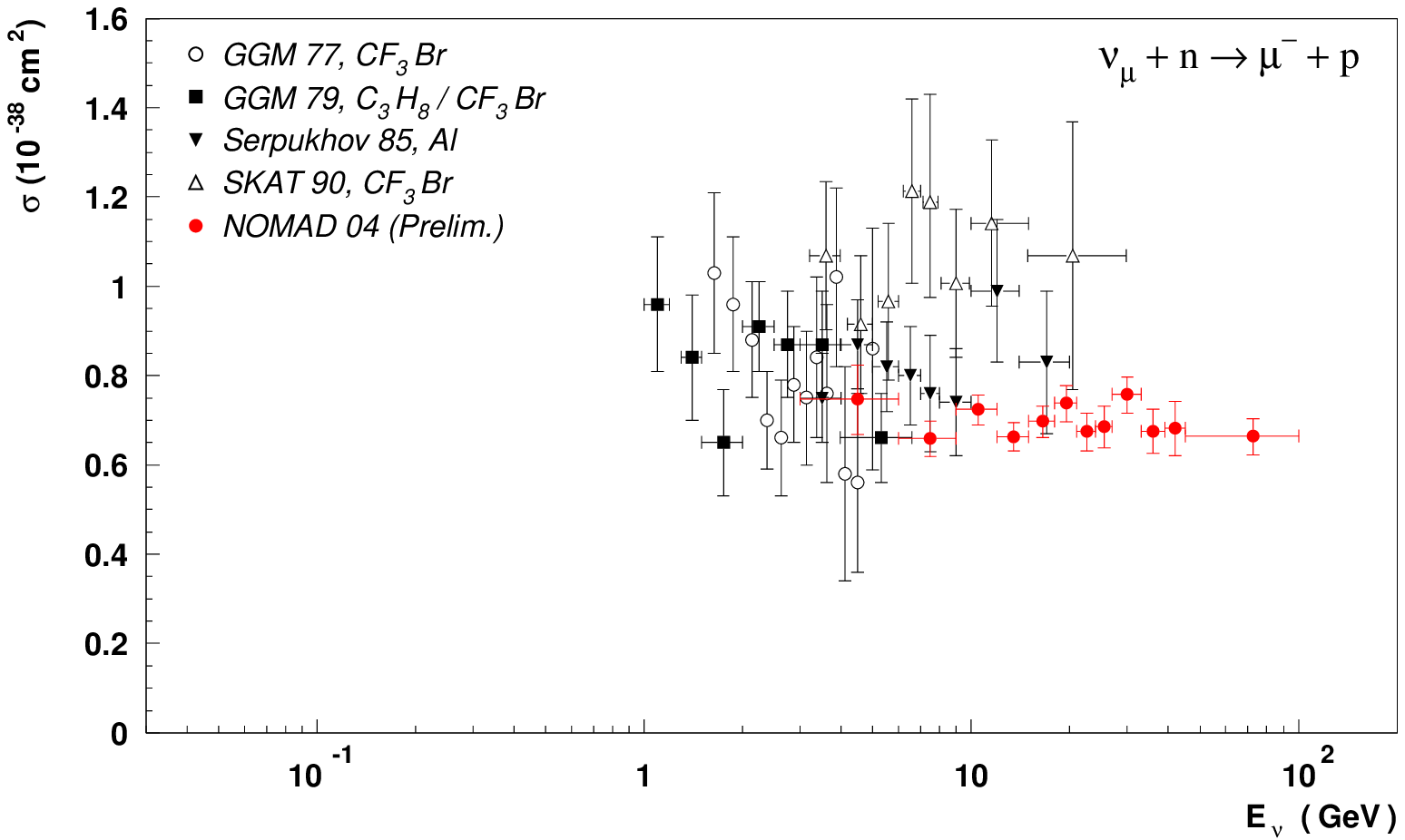}
\caption{Measurement of the $\nu_{\mu}$ quasi-elastic cross-section 
on carbon in NOMAD (full red circles) compared with all the available 
data. The upper plot shows results on deuterium 
targets, while the lower panel results on heavier targets.}
\label{fig:qelxsec}
\end{figure}

%The physics potential of the NOMAD data samples is  
%extremely rich and several physics topics are currently 
%under study. A brief summary of the main measurements is  
%given in the following sections.  

\subsection{Determination of $\sin^{2} \theta_{W}$} 
\label{sec:sin2thw}

A precise determination of the weak mixing angle from $\nu N$ 
Deep Inelastic Scattering (DIS) provides a {\em direct} measurement 
of neutrino couplings to $Z^{0}$, allowing a precision test of 
Standard Model at a different scale with respect to LEP/SLD. 
The interest of such measurement increased after the publication of
the final results on $\sin^{2}\theta_{W}$ by the NuTeV
collaboration~\cite{nutev}, reporting a deviation of about $3\sigma$
with respect to the Standard Model predictions~\cite{discussions}. 

The NOMAD experiment is extracting the value of $\sin^{2}\theta_{W}$ 
from the ratio of Neutral (NC) to Charged Current $\nu$ interactions: 
$$ 
{\cal R}_{\nu} = \frac{\sigma_{NC}}{\sigma_{CC}} 
$$ 
To reduce systematic uncertainties, a simultaneus fit 
to both ${\cal R}_{\nu}$ and the CC differential cross-section 
$d\sigma^{2}_{CC}/dx dy$ is performed. The analysis consists in 
classifying the events as NC and CC, providing the experimental 
ratio ${\cal R}_{\nu}^{\rm exp}$. The knowledge of the 
detection efficiencies and of the experimental resolutions 
provides then the cross-section values. 

The identification of CC interactions is based upon two {\em independent} 
criteria. First, events containing tracks (of any charge) matched to 
segments in the external muon chambers are flagged as $\nu_{\mu}$ CC. A kinematic 
tagging of the leading lepton is then applied to all events failing the 
previous requirement. This second tagging is provided by a multidimensional 
likelihood function~\cite{nmnt} and is applied to negative tracks only.  
The procedure provides two substantial advantages. A CC event failing any of the 
two complemetary criteria has a second possibility to be tagged as CC, thus
resulting in a reduction of the corresponding systematic uncertainty by one 
order of magnitude. In addition, the kinematic tagging is very efficient 
on electrons from $\nu_{e}$ CC events which are mostly identified 
without the use of any specific detector requirement. This in turn 
reduces systematic uncertainties from the $\nu_{e}$ beam content to a 
negligible level. The fraction of unidentified $\nu_{\mu}$ CC events 
is about 2\%. 

In principle, the theoretical model enters both in the Monte Carlo 
simulations used for the efficiency corrections and in the final fit 
to extract $\sin^{2}\theta_{W}$. However, a refined Data Simulator 
technique~\cite{nmnt} is used to extract all efficiencies from 
the data themselves, for both NC (with a NC-simulator) and CC (with a 
CC-simulator), through the expression 
$\epsilon = \epsilon_{\rm MC} \times \epsilon_{\rm DS} / \epsilon_{\rm MCS}$, 
where $\epsilon_{\rm MC}$, $\epsilon_{\rm DS}$ and $\epsilon_{\rm MCS}$ 
are the efficiencies extracted from the Monte Carlo, the Data Simulator and the 
Monte Carlo Simulator. The model largely cancels in the ratio 
$\epsilon_{\rm MC} / \epsilon_{\rm MCS}$. As a result, experimental efficiencies 
are stable against changes in the simulations.  

\begin{table}
\begin{center}
\begin{tabular}{l|c}
Source of uncertainty & $\delta {\cal R}_{\nu}/{\cal R}_{\nu}$ \\ \hline \hline
Statistics  & 0.00207 \\ \hline
%NC trigger efficiency  &  \\
%Charged multiplicity  &  \\
%Energy scale &  \\
%Muon ID &  \\
%$\mu$ acceptance  &  \\
%Hadron decays (NC) &   \\
%Kinematic selection &   \\
%Integral $\nu_{e}$ flux  &   \\ 
%Integral $\bar{\nu}_{\mu}$ flux  &   \\ 
%Integral $\bar{\nu}_{e}$ flux  &   \\ \hline
Experimental systematics  & 0.00194 \\ \hline
%Structure functions (LT)  &  \\
%High Twists (HT)  &  \\
%Strange sea  &  \\
%Charm mass  &  \\
%Nuclear effects  &  \\
%Non-isoscalarity  &  \\
%EW corrections  &  \\ \hline
Model systematics & 0.00181 \\ \hline\hline
TOTAL  & 0.00336 \\ \hline\hline
\end{tabular}
\label{tab:errors}
\caption{Summary of the relative uncertainties for the NOMAD measurement of the 
NC to CC ratio ${\cal R}_{\nu}$. The numbers refer only to the main data sample 
from the DC target (carbon).}
\end{center}
\end{table}

A preliminary estimate of the NOMAD sensitivity to ${\cal R}_{\nu}$ 
is shown in Table~\ref{tab:errors}. The dominant constributions 
to the systematic uncertainty are the determination of the NC 
trigger efficiency and the knowledge of the strange sea 
distribution. The corresponding precision on the 
$\sin^{2} \theta_{W}$ extraction is $\sim 1\%$, comparable 
to NuTeV~\cite{nutev}, E158~\cite{e158} and APV~\cite{apv}. It must be 
noted the addition of the sample of events from the front coil of the 
NOMAD magnet would reduce both the statistical uncertainty 
and the experimental systematics since the latter is 
defined by the size of the control samples from data.

The analysis is "blind" and the value of $\sin^{2}\theta_{W}$ 
will be available only after all the systematic uncertainties 
are finalized.

\subsection{Quasi-elastic $\nu_{\mu}$C cross-section}
\label{sec:qel} 

The reconstruction and identification of the recoiling proton 
track allowed a measurement of the quasi-elastic cross-section 
$\nu_{\mu} n \rightarrow \mu^{-} p$ on carbon in NOMAD.  
A kinematic analysis based upon a three-dimensional 
likelihood function is used to reject backgrounds from 
DIS and resonance production. Overall, 8192 events are selected 
with a signal efficiency of 25\% and a purity of 71\%.  
The measurement is performed in the energy range 
$3<E_{\nu}<100$ GeV. 

In order to reduce systematic uncertainties, NOMAD is 
measuring the {\em ratios} of quasi-elastic cross-section 
with respect to two independent processes: DIS 
($W^{2}>4$ GeV$^2$) and Inverse Muon Decay (IMD) 
$\nu_{\mu} e^{-} \rightarrow \mu^{-} \nu_{e}$. 
The absolute normalization is provided by 
the world average cross-section on isoscalar target 
for DIS and by the theoretical cross-section for IMD. 
Both measurements are consistent. Figure~\ref{fig:qelxsec}
shows a comparison of the preliminary NOMAD results  
on carbon with a compilation of existing data.

\subsection{Cross-sections and structure functions}
\label{sec:xsec} 

The present knowledge of neutrino cross-sections is rather 
nonuniform. In the region $E_{\nu}>30$ GeV, where data from the large 
massive calorimeters (CCFR, NuTeV) are available, the uncertainty is 
about 2\%. This increases to about 20\% at lower energies, due 
to the limited statistics of bubble chamber experiments. 
Measurements of both the total $\sigma_{CC}^{\rm tot}$ 
and the differential $d\sigma^{2}_{CC}/dx dy$ cross-sections for $\nu_{\mu}$ 
are performed in NOMAD, as part of the $\sin^{2}\theta_{W}$ analysis. 

%Two independent flux predictions are used. The first method~\cite{beam} 
%is based upon a detailed simulation of the beam line, including 
%constraints from SPY data on $\pi, K$ production in p-Be interactions.

\begin{figure}[h]
\epsfxsize220pt
\figurebox{120pt}{160pt}{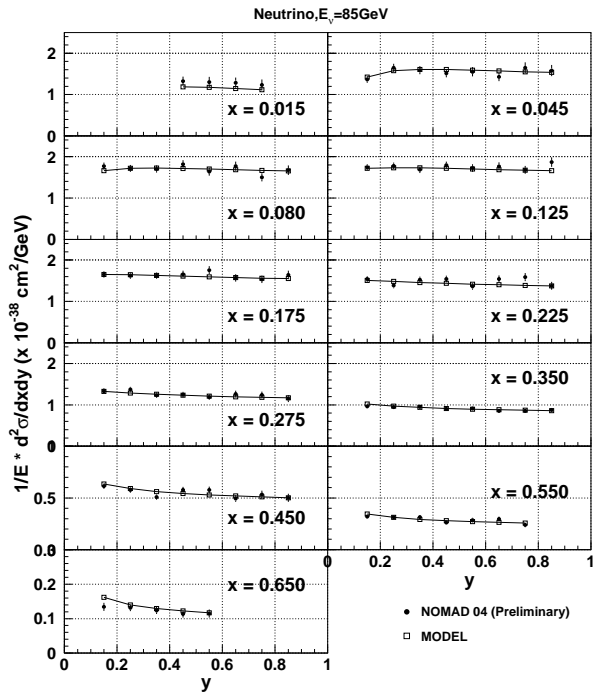}
\caption{Results from the NOMAD measurement of the $\nu_{\mu}$ 
differential cross-section $d\sigma^{2}_{CC}/dx dy$ on carbon at $E_{\nu}=85$ GeV 
(full circles). The open squares show the NNLO model predictions (Sec.~\ref{sec:model}),  
based upon the use of charged lepton scattering data only.}   
\label{fig:diffxsec}
\end{figure}

The absolute normalization is obtained from the world average 
cross-section on isoscalar target in the energy range $40<E_{\nu}<300$ GeV.  
A comparison of the measured differential cross-section on carbon 
with the model predictions (Sec.~\ref{sec:model}) is shown in Figure~\ref{fig:diffxsec}. 

The NOMAD cross-section data are also used to extract the structure 
functions $F_{2}$ and $xF_{3}$. The average $Q^2$ is about 13 GeV$^2$, with 
events extending to the few GeV$^2$ region. In addition, it is possible to 
study nuclear effects in neutrino structure functions by comparing the 
results with C, Al and Fe targets.

\subsection{Charm production and strange sea}
\label{sec:charm} 

The sample of events ($1.2\times10^{7}$ $\nu_{\mu}$ CC) originating in the 
forward iron calorimeter is used to measure the charm dimuon production 
in $\nu$ interactions. A total of $12767\pm 205$ identified 
charm events (after background subtraction) is selected, providing 
the single largest neutrino sample available.  

The value of the charm quark mass $m_c$, the strange sea suppression 
factor $\eta_{s}$ and the semileptonic branching ratio $B_{\mu}$ 
are extracted from a fit to kinematic distributions obtained from 
the {\em ratio} of dimuons to single muon events. Many systematic uncertainties 
cancel in such ratio. In addition, the energy spectrum of the NOMAD 
flux~\cite{beam} provides a good sensitivity to the charm threshold. 

\subsection{Modelling} 
\label{sec:model} 

The determination of $\sin^{2}\theta_{W}$ from ${\cal R}_{\nu}$ required the development 
of a refined modelling of neutrino interactions. A NNLO QCD 
calculation~\cite{alekhin} is performed to obtain the neutrino-nucleon structure functions. 
Dedicated fits to available data from charged lepton scattering on p and D (BCDMS, E665, HERA, JLab, NMC, SLAC) 
and from (anti)neutrino scattering (NOMAD, NuTeV), are performed to extract parton 
distributions, high twist terms and the corresponding uncertainties. %~\cite{newpdf}. 
 
A model of nuclear effects on structure functions is included~\cite{nucleff}. 
This has been defined on the basis of charged lepton nuclear data and checked 
against cross-section data. It includes shadowing, off-shell, Fermi motion and 
binding energy, pion contributions and non-isoscalarity corrections. 
 
A new evaluation of electroweak radiative corrections to structure 
functions is also used~\cite{ewcorr}. Results from an 
independent calculation~\cite{ewcorr2} allow a cross-check of 
systematic uncertainties.

A detailed tuning of fragmentation parameters is extracted with  
the analysis of individual tracks in the hadronic system reconstructed 
in $\nu_{\mu}$ CC data. 

The charm production parameters are determined primarily from the analysis 
of NOMAD dimuon data in FCAL (Sec.~\ref{sec:charm}). The mass of the 
charm quark ($\overline{MS}$ scheme) 
is further constrained by data from production thresold in $e^+e^-$ 
collisions~\cite{epem}. A precise parameterization of the strange sea 
distribution is obtained by including NOMAD dimuon data in the 
global fits to extract parton distributions. %~\cite{newpdf}.  

\end{document}